\documentstyle[twocolumn,aps,prl,epsf,floats]{revtex}
\begin{document} 
\twocolumn[\hsize\textwidth\columnwidth\hsize\csname @twocolumnfalse\endcsname
\title {
On the behavior of a 2D Heisenberg antiferromagnet at very low temperatures}
\author{Andrey V. Chubukov$^1$ and Oleg A. Starykh$^2$}
\address{
$^1$ Department of Physics, University of Wisconsin, Madison, WI 53706\\
$^2$ Department of Physics, University of Florida, Gainesville, FL 32611-8440}
\date{February 15, 1998}
\maketitle
\begin{abstract}
We present an analytical result for  
the ratio of the physical correlation length in a 2D 
Heisenberg antiferromagnet on
a square lattice, and the one which is actually computed in
numerical simulations. This last correlation length is
deduced from the second moment of the structure
factor at the antiferromagnetic momentum $Q$. 
We show that the ratio is very close to
one in agreement with previously obtained numerical 
result of the $1/N$ expansion.
\end{abstract}
]
The two-dimensional Heisenberg antiferromagnet on a square lattice 
is one of the most extensively
studied systems in condensed-matter physics. The interest to this model 
is two-fold. On one hand, Heisenberg antiferromagnet models a
large number of real materials including parent compounds of high-$T_c$
superconductors. On the other hand, its low-energy physics is adequately
described by a field-theoretical $\sigma-$model thus allowing one to 
find similarities between condensed-matter physics and field theory.

The low-temperature behavior of the Heisenberg antiferromagnet is understood
in great detail. For short-range interaction, the ground state is ordered
unless one fine tunes the couplings between nearest and further neighbors.
The ordered ground state is characterized by a sublattice order parameter,
$N_0$, spin stiffness, $\rho_s$, and
transverse susceptibility, $\chi_{\perp} = c^{-2} \rho_s$, where 
$c$ is the spin-wave velocity.
 At any finite temperature, however,
the system is disordered due to thermal fluctuations. The disordering 
means that equal-time spin-spin correlation function decays
exponentially with the distance, as $e^{-r/\xi}$. The 
length scale $\xi$  is the physical spin correlation length.
Various  approaches to $2D$ antiferromagnets all predict that 
in the renormalized-classical region ($T \ll \rho_s$) which we only consider
here, $\xi$ is
exponentially large in $T$ at low $T$ and behaves as
$\xi \sim \frac{c}{2\pi \rho_s} ~\exp{(\frac{2 \pi \rho_s}{T})}$.
Equal-time spin correlations at large distances can also be
described by a static structure
factor $S(k)$ for $k$ near the antiferromagnetic momentum $Q=(\pi,\pi)$.
In the disordered spin state, $S(Q)$ scales as $\xi^2$ and is therefore also
exponential in $T$.
 
The exponential temperature dependences of $\xi$ and of $S(Q)$ have 
been verified
in numerical simulations and by analyzing the neutron
scattering and NMR data for $La_2CuO_4$ and $Sr_2CuO_2Cl_2$. The accuracy
of numerical simulations is so high, however,
that one can not only check the
temperature dependences but also compare the absolute values of the structure
factor and the spin correlation length. 
There is, however, one subtlety in the numerical analyses - in numerical
simulations,  one measures not the physical
spin correlation length, but another length scale which is also exponentially
small in $T$, but generally differs from $\xi$ by a constant factor.   
Indeed, the physical spin correlation length can be extracted 
from the form of $S(k)$
simply because Fourier-transform of $S(k)$ yields the spin correlation function
in real space. However, one can easily make sure 
that the long-distance behavior of $S(r)$ is associated with 
the form of $S(k)$ along the {\it imaginary} $k$-axis. Moreover, the inverse
spin correlation length is the scale at which $S(k)$ has a pole
for imaginary $k$: $S^{-1} (k = i \xi^{-1}) =0$.
In numerical simulations, however, the structure factor is evaluated 
only for {\it real} values of momentum $k$.
Accordingly, a different definition of the correlation length is employed -
it is identified as a second moment of $S(k)$ for $k=Q$, i.e., as
${\tilde \xi}= (-S^{-1}(Q) dS(k)/dk^2|_{k \rightarrow
Q})^{1/2}$. For the  Lorentzian form of $S(k)$, 
$S(k) \propto ((Q-k)^2 + m_0^2)^{-1}$, both $\xi$ and ${\tilde \xi}$ are
equal to the  mean-field
spin excitation gap 
$m_0$ and are therefore undistinguishable. However, the $1/N$
calculations for the $O(N)$ $\sigma-$model rigorously demonstrated that
$S(k)$ has a Lorentzian form only in the limit $N \rightarrow \infty$ while
for arbitrary $N$, and, in particular, for physical $N=3$, $S(k)$ possesses a
much more complex dependence on $k$ which arizes due to temperature-independent
$1/N$ corrections. In this situation,
${\tilde \xi}$ and  the physical spin correlation length $\xi$ differ
by some constant factor. 
This factor appears to be a relevant one as several groups~ \cite{beard,troyer}
recently performed a detailed comparison of 
 $\tilde \xi$, calculated
numerically at very low $T$, with the exact expression for $\xi$
obtained some time ago by Hasenfratz and Niedermayer (see below). In this
analysis, they assumed that $\xi$ and ${\tilde \xi}$ are almost the same. 
{\it A priori}, however,
there are no reasons for such coincidence.

We start by quoting the seminal result \cite{HN} that
\begin{equation}
\xi^{-1}/m  = \left(\frac{8}{e}\right)^{1/(N-2)}~\frac{1}{\Gamma (1 + 1/(N-2))},
\label{aa}
\end{equation}
where $m$ is given by
\begin{equation}
m = \frac{T}{c}~\Big(\frac{2\pi \rho_s}{(N-2)T}\Big)^{\frac{1}{N-2}}
~e^{-\frac{2 \pi \rho_s}{(N-2)T}}.
\label{m}
\end{equation}
This result is based  on numerical results for $N=3$ and $N=4$~\cite{HN}
and on $1/N$ expansion for $O(N)$ $\sigma-$model~\cite{CSY,Camp}.
For $N=3$, this yields $\xi^{-1}/m = (8/e) \approx 2.94$.
No such exact expression, however, exists for $\tilde \xi$.
The $1/N$ expansion for $O(N)$ $\sigma$ model 
(which in the renormalized-classical region holds in powers of 
$1/(N-2)$~\cite{comm}) yields
\begin{equation}
{\tilde \xi}^{-1} = \xi^{-1} (1 + 0.003/(N-2)),
\label{xi1/N}
\end{equation}
where factor $0.003$ arises from numerical evaluation of some
complicated integrals.
A formal application of this result  to a physical case of 
$N=3$ yields almost identical values
for ${\tilde \xi}$ and $\xi$.  It is known, however, that a special care has to
be taken in this procedure 
as not all terms which appear to first order in $1/(N-2)$ 
actually contribute at $N=3$.
To illustrate this point,  consider large $N$ expansion for
$\xi^{-1}/m$. 
An analytical evaluation of the first $1/N$ correction yields
\begin{equation}
\xi^{-1}/m  = \left(1+ \frac{1}{N-2} \left(\log(8/e) + \gamma_E\right) \right)
\label{1/NN}
\end{equation}
where $\gamma_E$ is the Euler constant. Comparing this formula with the
exact result, Eq. (\ref{aa}), we notice that the Euler constant 
 accounts for the appearance of the $\Gamma$ function in (\ref{aa}):
$\Gamma(1 +1/(N-2)) =
1 -\gamma_E/(N-2) +O(1/(N-2)^2)$. As $\Gamma(2) = \Gamma(1) =1$, 
the term with the Euler constant does not contribute to $\xi^{-1}/m$ for the 
physical case of $N=3$.
The danger is that the same might also happen for
the rescaling factor between $\xi$ and $\tilde \xi$, i.e., that the ratio
obtained numerically to first order in $1/N$ may in fact contain the Euler
constant which would mask the actual value of the ratio.

In the present communication we address this issue.  We compute explicitly the
$T-$ independent  $1/N$ corrections to both correlation lengths and show that
the rescaling factor does not contain the Euler constant. This implies that
the $1/N$ result for the ratio is very likely to be trustable, and the
rescaling factor between two correlation length is very close to one.

As we already said,  in our consideration, we will use heavily 
results of the $1/N$ expansion for $O(N)$ $\sigma-$model~\cite{CSY,Camp}.
The point of departure for this analysis is the  mean-field theory
which is exact at $N=\infty$. At the mean-field level,
the static structure factor is given by
\begin{equation}
S(k) = \sum_{i=1}^{N} S_{i,i} (k) = \frac{T N^2_0 N}{\rho_s}~\frac{1}
{m^2_0 +(Q-k)^2}
\end{equation}
and $m_0 = (T/c) e^{-2\pi \rho_s/(NT)}$. 
Consider now finite $N$. 
It has been shown in ~\cite{CSY,Camp} that there exist
 two different types of $1/(N-2)$ corrections: singular ones which contain 
$\log(T/m_0)$ and
$\log(\log(T/m_0))$, and regular ones, which at small $T$ account for the
$T$-independent renormalizations of $S(k)$ and $m$. 
There is a large amount of confidence that 
logarithmical and double logarithmical perturbation series are 
geometrical and therefore can be simply exponentiated~\cite{comm3}.
Regular $1/(N-2)$ corrections require a special care, as we just demonstrated.
 Collecting both singular
and regular $1/N$ corrections and exponentiating the singular ones, one 
obtains~\cite{CSY}
\begin{equation}
S (k) = 2\pi N^2_0 \frac{N}{N-2} ~\Big(\frac{(N-2) T}{2\pi\rho_s}\Big)^
{\frac{N-1}{N-2}} P(k)
\label{q}
\end{equation}
where for $k$ comparable to the inverse correlation length
\begin{equation}
P(k) = \frac{1}{Z m^2 +(Q-k)^2 + \Sigma(k)} .
\label{Q}
\end{equation}
Here $m$ is given by (\ref{m})  
and $Z$ and $\Sigma(k) \propto (Q-k)^2$ account for a temperature independent
$1/(N-2)$ corrections.
For $N=3$, the functional forms of $S(k)$ and $m$ obtained this way
fully agree with the ones obtained in the perturbative RG approach \cite{CHN}.

 The expressions for $Z$ and $\Sigma (k)$ have been obtained in~\cite{CSY} 
 but not explicitly presented in that paper. Here we list the catalog of the
results which we will need:
\begin{eqnarray}
Z=&&1+ \frac{2}{N} \Big(2 \log{2} -1 - 
\nonumber\\
&&
3 \int_0^{\infty}dx \log \log{\frac{x + \sqrt{x^2 +4}}{2}}
~\frac{x}{(x^2 +1)^2}\Big);\nonumber\\
\Sigma (k \rightarrow Q) =&& - \frac{4 (Q-k)^2}{N} \int_0^{\infty}dx 
 \log \log{\frac{x + \sqrt{x^2 +4}}{2}}
\nonumber\\
&&\times ~\frac{x (7x^2 -2)}{(x^2 +1)^4};\nonumber\\
\Sigma (k =im) =&& - \frac{m^2}{N} \int_0^{\infty}dx 
 \log \log{\frac{x + \sqrt{x^2 +4}}{2}} 
\nonumber\\
&&\times~\left(\frac{(x-\sqrt{x^2
+4})^2}{\sqrt{x^2 +4}} - 6\frac{x}{(x^2+1)^2}\right).
\label{int}
\end{eqnarray}
For the two correlation lengths we then obtain
\begin{eqnarray}
\xi^{-2} =&& m^2 Z (1 + \frac{\Sigma (k =im)}{m^2});\nonumber\\
{\tilde \xi}^{-2} =&& m^2 Z (1 - \frac{\Sigma(k)}{(Q-k)^2})|_{k \rightarrow Q}.
\end{eqnarray}
Performing simple manipulations, we find
\begin{equation}
\xi^{-1} = m \left(1 + \frac{1}{N} \left(2\log{2} -1 -A\right)
\right),
\label{zz}
\end{equation}
where 
\begin{equation}
A =\int_0^{\infty}dx 
 \log \log{\frac{x + \sqrt{x^2 +4}}{2}} ~\left(\frac{(x-\sqrt{x^2
+4})^2}{2\sqrt{x^2 +4}}\right).
\end{equation}
Introducing $t = \log{\frac{x + \sqrt{x^2 +4}}{2}}$ and integrating by parts,
 we immediately obtain $A = -(\gamma_E + \log{2})$. 
Substituting this result into (\ref{zz})
we recover Eq. (\ref{1/NN}). For the ratio of $\xi$ and ${\tilde \xi}$, the
same manipulations yield a more complex expression
\begin{equation}
{\tilde \xi} = \xi \left(1 + \frac{1}{N} \left(\gamma_E + I\right)\right)   
\label{I}
\end{equation}
where
$I = 3I_2 -14I_3 +18I_4$, and
\begin{equation}
I_n = \int_0^{\infty} dt \frac{\sinh{t} \log{t}}{(2\cosh{t}-1)^n}
\label{ii}
\end{equation} 
Notice that all integrals $I_n$ are convergent. Physically, this implies that 
$T-$independent factors in both correlation lengths are determined by system
behavior at energy
scales which are much smaller than the upper cutoff. At this scales,
the behavior is universal  and therefore overall factors in $\xi$ and
${\tilde \xi}$ are also universal numbers.
Note by passing that similar integrals appear in the 
calculations of the d.c. Hall conductivity 
near fractional quantum Hall critical point~\cite{subir}. 

To evaluate the integrals (\ref{ii}), 
we introduce auxiliary function $\Phi_n (t) =    
\sinh{t} \log^2(-t)/(2\cosh{t}-1)^n$ and integrate $\Phi_n$ over a contour
which consists of a circle of an infinite radius and a cut along positive real
$t$. The contour integral yields $-4\pi i I_n$ and simultaneously it is equal
to the sum of the residues (modulo $2\pi i$) 
of the poles along imaginary $t-$axis. Performing
calculations 
and making use of the summation formula
\begin{eqnarray}
&&\sum_{n=0}^{\infty} \frac{\log(2\pi n + \pi(1-a))}{2n +(1-a)} -
\frac{\log(2\pi n + \pi(1+a))}{2n +(1+a)} =
\nonumber\\
&& \frac{\pi}{2} (\log{\pi} -
\gamma_E)~\tan \frac{\pi a}{2} - \int_0^{\infty} du \frac{\sinh{ua}}{\sinh{u}}
\log{u},
\end{eqnarray} 
we can explicitly pull out the Euler constant from the integrals
\begin{eqnarray}
I_2 &=& -\frac{\gamma_E}{2} - \frac{{\tilde I}_2}{2}, 
~I_3 =-\frac{\gamma_E}{4} - \frac{{\tilde I}_2}{12} + \frac{{\tilde
I}_3}{12},\nonumber\\ 
I_4 &=&-\frac{\gamma_E}{6} - \frac{{\tilde I}_2}{18} + \frac{{\tilde I}_3}{36} -
\frac{{\tilde I}_4}{36}
\label{I_n}
\end{eqnarray}
where
\begin{eqnarray}
{\tilde I}_2 &=& \int_0^\infty \frac{\log{x}}{\sinh{\pi x}}~\frac{\sinh{2 \pi
x/3}}{\sin{2\pi/3}},\nonumber\\
{\tilde I}_3 &=& \int_0^\infty \frac{x \log{x}}{\sinh{\pi x}}~\frac{\cosh{2 \pi
x/3}}{\cos{2\pi/3}},\nonumber\\
{\tilde I}_4 &=& \int_0^\infty \frac{x^2 \log{x}}{\sinh{\pi x}}~\frac{\sinh{2 
\pi x/3}}{\sin{2\pi/3}}
\end{eqnarray}
The analytical expressions for the 
integrals ${\tilde I}_{3,4}$ have been obtained  very recently~\cite{Adamchik},
whereas ${\tilde I}_2$  was known for some time~\cite{grad}.
It turns out that these 
integrals can be expressed in terms of the derivatives of the Hurvitz Zeta 
function 
$d \zeta(x, \alpha)/dx$ at $x=0,-1,-2$ and $\alpha=1/6,1/3,1/2,2/3,5/6,1$. 
For $x=0$, $d \zeta(x, \alpha)/dx = 
\log(\Gamma(\alpha)/\sqrt{2\pi})$. By analogy, we can introduce 
generalized Gamma functions via
$d \zeta(x, \alpha)/dx |_{x=-n} =
 \log(\Gamma_{-n}(\alpha)/\sqrt{2\pi}) $. Using this definition, we obtain
a symbolic representation
\begin{eqnarray}
{\tilde I}_2 =&& -\log R_0; ~{\tilde I}_3 = 6 \log R_{-1} -3 \log 6; 
\nonumber\\
{\tilde I}_4 =&& 36 \log R_{-2} + 3 \log 6
\end{eqnarray}
where
\begin{equation}
R_{-s} = \frac{\Gamma_{-s} (1/3) \Gamma_{-s} (1/2)}{\Gamma_{-s} (5/6)
\Gamma_{-s}(1)}~\left(\frac{\Gamma_{-s} (7/6) \Gamma_{-s} (1)}{\Gamma_{-s} (2/3)
\Gamma_{-s}(1/2)}\right)^{(-1)^s}
\end{equation}

The expression for $R_0$ and $R_{-1}$ can be reduced to an explicit 
 closed-form expression
containing, e.g., logarithms of Gamma-functions, but no Euler
constant~\cite{comm2}. 
No closed-form
expression in terms of tabulated functions is known for $R_{-2}$. 
It is however quite reasonable to assume that  it also does not 
contain Euler constant.

Assembling now all contributions to $I$ and substituting the result into
(\ref{I}) we find that the Euler constant is cancelled out. 
 The rest is combined into 
\begin{equation}
{\tilde \xi} = \xi \left(1 + \frac{1}{2(N-2)} \log [6 R^{8/3}_0 R^{-8}_{-1}
R^{-36}_{-2}] + O\left(\frac{1}{(N-2)^2}\right)\right)   
\end{equation}
This expression is the central result of the paper.

The next issue is how to account for the higher-order terms in $1/(N-2)$. 
Here we use the same assumption as was proven to work for $\xi$, namely that 
 after the Euler constant is subtracted, the rest of the
regular  $O(1/(N-2))$ correction can be exponentiated.
Using this assumption,  we finally obtain 
\begin{equation}
{\tilde \xi}^2 = \xi^2 ~(6 R^{8/3}_0 R^{-8}_{-1} R^{-36}_{-2})^{1/(N-2)}   
\end{equation}
For $N=3$ this yields ${\tilde \xi}^2/\xi^2 =0.993$, i.e. the ratio is indeed
very close to one.  This result may sound intuitively obvious, but 
is not based on any apparent physical reasons
and therefore had to be verified by  explicit calculations. This is
what we did.

The extreme closeness of the ratio
${\tilde \xi}/\xi$ to $1$ is consistent with recent claims 
that at $T \rightarrow 0$,  the numerically computed spin correlation length~
\cite{beard,troyer}
approaches the Hasenfratz-Niedermayer result, Eq.(\ref{aa}).

Using our expressions for $Z$ and $\xi$, we can also compute the overall
factor for the structure factor $S(Q)$ which is also studied in numerical
simulations. 
 The numerical evaluation of the first $1/(N-2)$ correction yields
$P(Q) \equiv 1/(Zm^2) =\xi^2 (1 +0.188/(N-2))$. Exponentiating this result,
 one obtains $P(Q)/\xi^2 \approx 1.2$. Numerical 
simulations \cite{troyer}, on
the other hand, reported that the actual rescaling factor is more than three
times larger than this number.  We performed analytical calculations along the
same lines as above and obtained 
\begin{equation}
P(Q) = \xi^2 [1 + \frac{1}{N-2} \left( 2\gamma_E - \log{2} + 6 I_2\right)]
\end{equation}
where $I_2$ is given by (\ref{I_n}). Substituting the result for $I_2$ into
this expression, we obtain
\begin{equation}
P(Q) = \xi^2 [1 + \frac{1}{N-2} \left( -\gamma_E +
3\log(R_0) - \log{2}\right)]
\label{P}
\end{equation}
This result coincides with the one obtained earlier by Campostrini and Rossi
\cite{Camp}, and cited previously in \cite{Sandvik}.
We see now that Euler constant is present in the perturbation series,
 i.e. one cannot simply exponentiate the lowest-order result.
Using the same procedure as before, i.e., 
 treating $\gamma_E$ as coming from the expansion of
 $\Gamma(1 + 1/(N-2))$, and exponentiating the rest of (\ref{P}), we obtain
\begin{equation}
P(Q) = 2^{1/(2-N)}~\Gamma(\frac{N-1}{N-2}) \left(\frac{\Gamma (1/3) 
\Gamma (7/6)}{\Gamma (5/6)
\Gamma(2/3)}\right)^{3/(N-2)} \xi^2
\end{equation}
For $N=3$, this yields $P(Q) \xi^{-2} =2.149$ which is about two times larger 
than the 
result obtained by formally exponentiating the whole $1/(N-2)$ correction.
Still, however,  this result does not
fully agree with quantum Monte Carlo simulations at low $T$ which 
reported $P(Q)\xi^{-2} \approx 4$ for both $S=1/2$~\cite{troyer} and
$S=1$~\cite{s=1}. The series expansion results~\cite{elstner}
reported somewhat smaller  $P(Q)\xi^{-2} \approx 3.2$ for $S=1/2$.
The reason for the discrepancy is not clear to us. 
Possibly, numerical simulations for $S(Q)$ were performed 
not deep inside the asymptotic scaling regime at $T \rightarrow 0$. Another
possibility is that something may be wrong with the exponentiation of
the first $1/N$ correction to $S(Q)$, though this is unlikely in view of the
results for $\xi$. 

We would like to thank V. Adamchik, S. Sachdev, M. Troyer,
R. Wickham and  M. Zhitomirsky for fruitful discussions.


\begin{references}

\bibitem{beard} B. B. Beard, R. J. Birgeneau, M. Greven, and U. J. Wiese,
cond-mat/9709110; to appear in \prl {\bf 80} (1998).

\bibitem{troyer} J. K. Kim and M. Troyer, cond-mat/9709333; to appear
in \prl {\bf 80} (1998).

\bibitem{HN} P. Hasenfratz, M. Maggiore, and F. Niedermayer, Phys. Lett. B
{\bf 245}, 522 (1990); P. Hasenfratz and F. Niedermayer, {\it{ibid.}}
{\bf 245}, 529 (1990).

\bibitem{CSY} A. V. Chubukov, S. Sachdev, and J. Ye, \prb {\bf 49}, 11919
(1994).

\bibitem{Camp} M. Campostrini and P. Rossi, Rivista del Nuovo Cimento {\bf 16},
No.6, 1 (1993).

\bibitem{CHN} S. Chakravarty, B. I. Halperin, and D. R. Nelson, \prb {\bf 39},
2344 (1989).

\bibitem{comm} The transformation $N \rightarrow N-2$ is due to the fact that
in the renormalized-classical region, the low-temperature behavior of the
system  is governed by the interaction between would-be 
Goldstone modes, and a number of them interacting with the given one
is $N-2$.

\bibitem{comm3} The confidence is based on the results of the $1/N$ expansion
to second order in $1/N$ and on the full agreement, for the physical $N=3$,
 between the functional forms of observables, obtained by exponentiation of
 the leading $1/N$ corrections, and the ones obtained by the perturbative RG
treatment~\protect\cite{CHN}.
 
\bibitem{subir} S. Sachdev, cond-mat/9709243.

\bibitem{Adamchik} V. S. Adamchik, {\it{A class of logarithmic integrals}},
in  Proceedings of ISSAC'97, Maui, USA, ACM Press, New York, 1997.

\bibitem{grad} I. S. Gradshtein and I. M. Ryzhik, {\it{Tables of Integrals, 
Sums, and Products}}, Academic Press, New York, 1980.

\bibitem{comm2}
The closed forms for $R_{0}$ and $R_{-1}$ are: 
$R_0=\frac{\Gamma(1/3)\Gamma(7/6)}{\Gamma(2/3) \Gamma(5/6)}$.
$R_{-1}=e^{-1/3} 2^{1/18} 3^{1/8} A_G^4$, where $A_G$ is Glaisher's 
constant, $\log(A_G)=1/12 - \zeta^{\prime}(-1)$.

\bibitem{Sandvik} A. W. Sandvik, A. V. Chubukov, S. Sachdev, \prb {\bf 51}, 
16483 (1995). Eq. (5) in this paper contains misprints.

\bibitem{s=1} K. Harada. M. Troyer, and N. Kawashima, cond-mat/9712292.

\bibitem{elstner} N. Elstner, A. Sokol, R. R. P. Singh, M. Greven, 
R. J. Birgeneau, \prl {\bf 75}, 938 (1995).

\end{references}
\end{document}